\documentclass[onecolumn]{aastex631}

\begin{document}

\title{Widespread Hot Molecular Gas Heated by Shear-induced Turbulence in the Galactic Center}

\author[0000-0003-3520-6191]{Juan Li}\thanks{lijuan@shao.ac.cn}
\affiliation{Shanghai Astronomical Observatory, Chinese Academy of Sciences, No. 80 Nandan Road, Shanghai, 200030, P. R. China}
\affiliation{State Key Laboratory of Radio Astronomy and Technology, A20 Datun Road, Chaoyang District, Beijing, 100101, P. R. China}

\author[0000-0001-6106-1171]{Junzhi Wang}\thanks{junzhiwang@gxu.edu.cn}
\affiliation{Guangxi Key Laboratory for Relativistic Astrophysics, Department of Physics, Guangxi University, Nanning 530004, P. R. China}

\author{Zhiqiang Shen}
\affiliation{Shanghai Astronomical Observatory, Chinese Academy of Sciences, No. 80 Nandan Road, Shanghai, 200030, P. R. China}
\affiliation{State Key Laboratory of Radio Astronomy and Technology, A20 Datun Road, Chaoyang District, Beijing, 100101, P. R. China}

\author{Alba Vidal-Garc{\'\i}a}
\affiliation{Observatorio Astron\'omico Nacional (OAN, IGN), C/ Alfonso XII, 3, 28014, Madrid, Spain}

\author[0000-0002-2243-6038]{Yuqiang Li}
\affiliation{Shanghai Astronomical Observatory, Chinese Academy of Sciences, No. 80 Nandan Road, Shanghai, 200030, P. R. China}
\affiliation{Korea Astronomy and Space Science Institute, No. 776, Daedeok-daero, Yuseong-gu, Daejeon, Republic of Korea}
\affiliation{School of Astronomy and Space Sciences, University of Chinese Academy of Sciences, No. 19A Yuquan Road, Beijing 100049, P. R. China}
\affiliation{State Key Laboratory of Radio Astronomy and Technology, A20 Datun Road, Chaoyang District, Beijing, 100101, P. R. China}

\author{Di Li}
\affiliation{New Cornerstone Science Laboratory, Department of Astronomy, Tsinghua University, Beijing 100084, P. R. China}
\affiliation{National Astronomical Observatories, Chinese Academy of Sciences, 20A Datun Road, Chaoyang District, Beijing 100012, P. R. China}

\author{Liubin Pan}
\affiliation{School of Physics and Astronomy, Sun Yat-sen University, Daxue Road, Zhuhai, Guangdong, 519082, P. R. China}

\author{Lei Huang}
\affiliation{Shanghai Astronomical Observatory, Chinese Academy of Sciences, No. 80 Nandan Road, Shanghai, 200030, P. R. China}
\affiliation{State Key Laboratory of Radio Astronomy and Technology, A20 Datun Road, Chaoyang District, Beijing, 100101, P. R. China}

\author{Fengyao Zhu}
\affiliation{Research Center for Intelligent Computing Platforms, Zhejiang Laboratory, Hangzhou 311100, P. R. China}

\author[0000-0001-9047-846X]{Siqi Zheng}
\affiliation{Shanghai Astronomical Observatory, Chinese Academy of Sciences, No. 80 Nandan Road, Shanghai, 200030, P. R. China}
\affiliation{School of Astronomy and Space Sciences, University of Chinese Academy of Sciences, No. 19A Yuquan Road, Beijing 100049, P. R. China}
\affiliation{I. Physikalisches Institut, Universit\"{a}t zu Köln, Z\"{u}lpicher Str. 77, 50937 K\"{o}ln, Germany}
\affiliation{State Key Laboratory of Radio Astronomy and Technology, A20 Datun Road, Chaoyang District, Beijing, 100101, P. R. China}

\author{Yiping Ao}
\affiliation{Purple Mountain Observatory, Chinese Academy of Sciences, Nanjing 210008, P. R. China}
\affiliation{State Key Laboratory of Radio Astronomy and Technology, A20 Datun Road, Chaoyang District, Beijing, 100101, P. R. China}

\author{Alvaro Sanchez-Monge}
\affiliation{Institut de Ci\`encies de l’Espai (ICE), CSIC, Campus UAB, Carrer de Can Magrans s/n, 08193 Bellaterra (Barcelona), Spain}
\affiliation{Institut d’Estudis Espacials de Catalunya (IEEC), 08860, Castelldefels (Barcelona), Spain}

\author{Zhiyu Zhang}
\affiliation{School of Astronomy and Space Science, Nanjing University, Nanjing 210093, P. R. China }
\affiliation{Key Laboratory of Modern Astronomy and Astrophysics (Nanjing University), Ministry of Education, Nanjing 210093, P. R. China}

\author[0000-0003-2619-9305]{Xing Lu}
\affiliation{Shanghai Astronomical Observatory, Chinese Academy of Sciences, No. 80 Nandan Road, Shanghai, 200030, P. R. China}
\affiliation{State Key Laboratory of Radio Astronomy and Technology, A20 Datun Road, Chaoyang District, Beijing, 100101, P. R. China}

\author[0000-0002-5286-2564]{Tie Liu}
\affiliation{Shanghai Astronomical Observatory, Chinese Academy of Sciences, No. 80 Nandan Road, Shanghai, 200030, P. R. China}
\affiliation{State Key Laboratory of Radio Astronomy and Technology, A20 Datun Road, Chaoyang District, Beijing, 100101, P. R. China}

\author{Xingwu Zheng}
\affiliation{School of Astronomy and Space Science, Nanjing University, Nanjing 210093, P. R. China }
\affiliation{Key Laboratory of Modern Astronomy and Astrophysics (Nanjing University), Ministry of Education, Nanjing 210093, P. R. China}

\begin{abstract}

We observed NH$_3$ metastable inversion lines from (3, 3) to (18, 18) toward G0.66-0.13 in the Galactic center with the Shanghai Tianma 65m radio telescope and Yebes 40 m telescope. Highly-excited lines of NH$_3$ (17, 17), (18, 18) were detected in emission for the first time in the interstellar medium, with upper energy levels up to 3100 K. Mapping observations reveal widespread hot molecular gas traced by NH$_3$ (13, 13) toward G0.66-0.13. The rotation temperatures of hot gas traced by NH$_3$ exceed 400 K, which amounts to five percent of the total NH$_3$ in the Galactic Center. Hot gas ($>$400 K) and warm gas (100-140 K) are found in distinct clumps, with the hot gas located at the interfacing regions between different warm clouds. The theory of intermittency in turbulence reproduces the complex temperature structure in the central molecular zone, especially the hot gas observed here. The results presented here demonstrate that turbulence heating dominates the heating of the molecular gas in the Central Molecular Zone, while the turbulence is induced by the shear-motion of molecular clouds under the gravitational potential of the nuclear star clusters and the supermassive black hole. Our results suggest that shear-induced turbulence heating could be a widespread factor influencing galactic evolution.

\end{abstract}


\section{Introduction} 
\label{introduction}

Located at 8.2 kiloparsecs from the Earth~\citep{gravity2019}, our Galactic Center serves as a unique laboratory for studying extreme astrophysical processes. The majority of the molecular gas in the GC is concentrated in a chain of Giant Molecular Clouds forming the so-called Central Molecular Zone (CMZ)~\citep{Henshaw2023}. This region accounts for nearly 5\% of the Galaxy's total molecular gas reservoir~\citep{Dahmen1998, Nakanishi2006}, and it hosts more than 80\% of the dense gas found in the Galaxy. The properties of molecular gas in the CMZ are much more extreme than in the Galactic disk. With the average density of $\sim10^4$ cm$^{-3}$, it is two orders of magnitude greater than typical disk values~\citep{Bally1987, Henshaw2023}. Line widths due to turbulent motions are also large, with full-width half-maximum (FWHM) ranging from 2.6 to 53.1 km s$^{-1}$~\citep{Henshaw2016}. Gas temperatures are higher than those in the disk as well ($T_{CMZ}$=50-400 K versus $T_{disk}$=10-20 K)~\citep{Huettemeister1993, Huettemeister1995, Rodriguez2001, Wilson2006, Ao2013, Mills2013, Ginsburg2016}. Previous molecular gas temperature measurements reveal complex temperature structures in the CMZ, with a prominent hot molecular gas component of temperatures above $350$~K~\citep{Huettemeister1995, Rodriguez2001, Wilson2006, Mills2013}. 

Ammonia (NH$_3$) is perhaps the most important tracer for the physical conditions of molecular clouds as it has a relatively low critical density (a few times 10$^3$ cm$^{-3}$)~\citep{Ho1983}. \citet{Mills2013} detected metastable transitions of NH$_3$ (8, 8) - (15, 15) toward several positions in the CMZ. They obtained rotational temperatures ranging from 350 K to 450 K from these metastable lines. Their observations reveal the widespread presence of hot molecular gas in the Galactic center. In M-0.02-0.07, the hot NH$_3$ is found to contribute $\sim$ 10\% of the cloud's total NH$_3$ column density, and the hot NH$_3$ seems to arise in gas which is extended or uniformly distributed on $\leq$10 arcsec scale. The nature of the hot gas component is unclear. Recently, using NH$_3$ metastable inversion lines of (8, 8)- (14, 14), \citet{Candelaria2023} reported detection of hot component with temperatures ranging from 210 - 580 K in the CMZ. \citet{Wilson2006} detected the (18,18) inversion line of ammonia, with a upper level energy of 3130 K above the ground state, in absorption towards Sagittarius B2(M) (hereafter Sgr B2(M)). Whether the (18, 18) ammonia absorption towards Sgr B2 comes from the extended low-density envelope surrounding Sgr B2(M) and (N), or much more compact, dense clouds associated with the hot cores of this star-forming region is unclear. 

The rapid cooling of hot molecular gas makes sustained temperatures of hundreds of Kelvins thermodynamically challenging. So far, how the hot gas is heated and sustained remains unclear. Possible mechanisms include turbulence~\citep{Ao2013, Ginsburg2016, Goicoechea2013, Immer2016}, X-ray heating~\citep{Maloney1996}, cosmic-ray heating~\citep{Goldsmith1978, Bradford2003, Ao2013}, UV heating through photodissociation regions \citep{Rodriguez2004}, HII region heating \citep{Vogel1987},  and large-scale instabilities~\citep{Kruijssen2015}. \citet{Ao2013} investigated heating mechanism in the Galactic center quantitatively. Their results show that the high temperatures found in the Galactic center  region may be caused by turbulent heating and/or cosmic-ray heating. \citet{Ginsburg2016} proposed that turbulent heating can readily explain the observed temperatures given the observed line widths. They argued that cosmic rays cannot explain the observed variation in gas temperatures. By comparing the temperature maps with the kinematic structure of M-0.02-0.07, a cloud in the Sgr A complex, \citet{Mills2013} hypothesized that the high temperatures may be due to the collision of different velocity components of the cloud. In addition, a hot CO component at high density was detected with {\it Herschel}, which was regarded to result from a combination of UV- and shock-driven heating~\citep{Goicoechea2013}. %Mapping observations of hot gas in the CMZ may shed light on this issue.

\begin{figure}
\begin{center}
\includegraphics[width=0.85\textwidth]{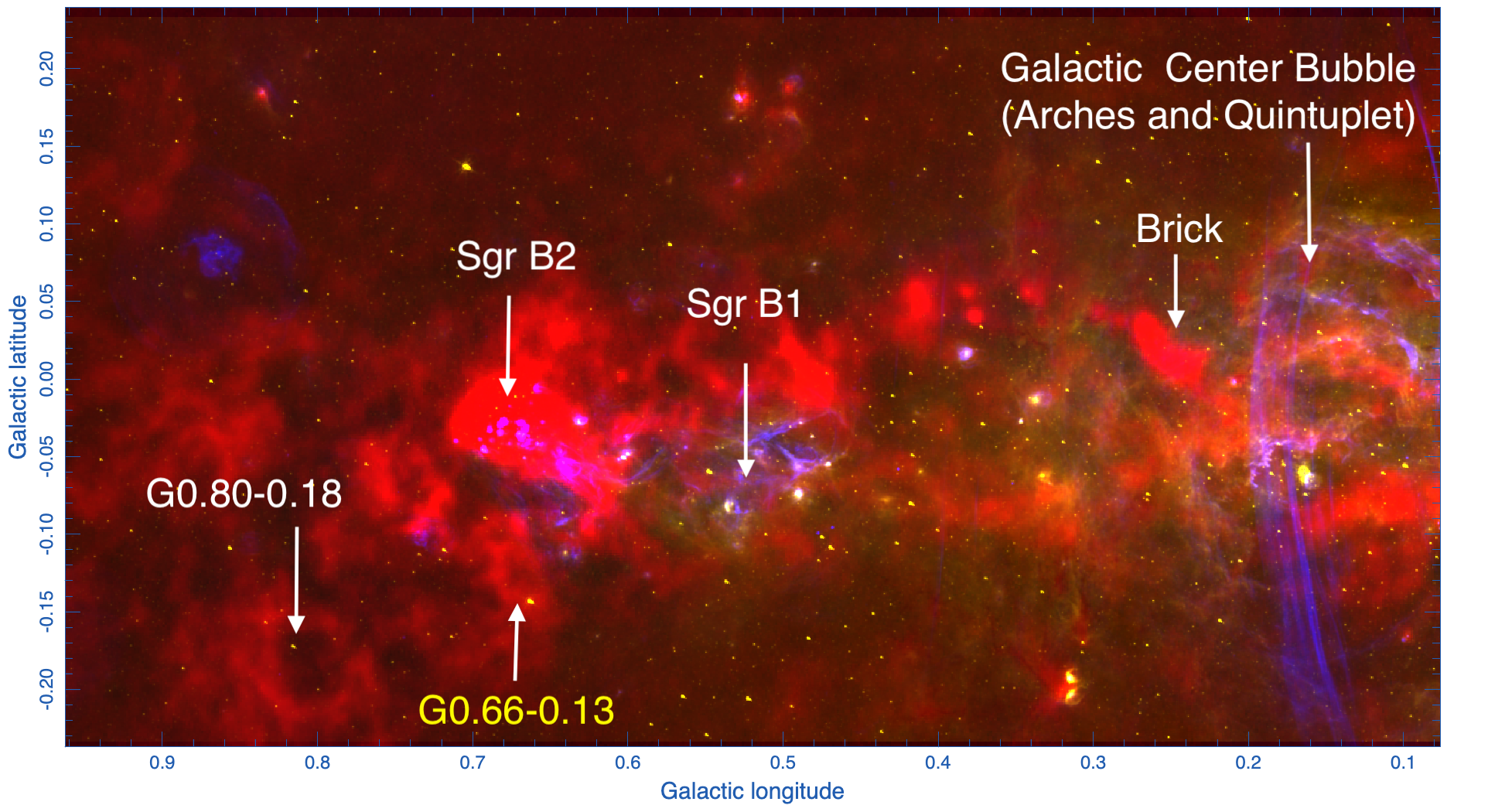}
\caption{The three-colour composite image of 8 µm emission from the {\it Spitzer} GLIMPSE survey (yellow; \citet{Churchwell2009}), 870 µm emission from the ATLASGAL survey (red; \citet{Schuller2009}), and 20 cm emission observed by MeerKAT (blue; \citet{Heywood2022}). A number of interesting features in the Galactic Center are overlaid. }
\label{fig:cmz} 
\end{center}
\end{figure}

Sgr B2, the most massive molecular cloud in the CMZ, is of particular interest to explore the physical environment of the GC. G0.66-0.13 is a quiescent molecular cloud located southeast of Sgr B2 (see Figure~\ref{fig:cmz} for the location of G0.66-0.13 in the CMZ). Bright SiO emission~\citep{Tsuboi2015, Armijos2020, Banda2023}, mid-J CO~\citep{Santa2021}, strong X-ray emission~\citep{Zhang2015}, and widespread complex organic molecules~\citep{Li2017,Li2020} have been detected toward G0.66-0.13. Shocks and enhanced cosmic-ray fluxes were thus believed to coexist in this region~\citep{Santa2021}. The absence of detectable star-formation signatures (e.g. Class 0/1 YSOs~\citep{An2011}, H$_2$O masers~\citep{Chambers2014}, and class II CH$_3$OH masers~\citep{Lu2019}) suggests that the physical conditions of this cloud are not dominated by stellar feedback. In this paper we present observing results of NH$_3$ (3, 3)- (18, 18) toward G0.66-0.13. Excitation temperatures were obtained and the origin of hot gas was discussed. In Section \ref{observations}, we describe the observations and data reduction. In Section \ref{results}, we present the observing result. We discuss the result in Section \ref{discussion} and then summarize our conclusions in Section \ref{conclusion}.

%__________________________________________________________________

\section{Observations and data reduction}
\label{observations}

The observational data used in this paper are based on observations of the the Shanghai Tianma 65m radio telescope (TMRT) in Shanghai, China and the Yebes 40m telescope in Guadalajara, Spain. The Yebes observations were made during the period of October, 2022 to December, 2022 (Project ID: 22B016, principle investigator (PI): J. Li) and June, 2024 (Project ID: 24A002, PI: J. Li), while the Tianma observations were made during September, 2023, to March, 2025. All the lines presented in this work are summarized in Table~\ref{tab:line}. 

\subsection{Yebes 40m observations}

We used the Q-band (7 mm) HEMT receiver developed within the Nanocosmos project~\citep{Tercero2021}. The Q-band receiver consists of two high electron mobility transistor cold amplifiers covering the 31.0-50.3 GHz range with horizontal and vertical polarizations. Eight 2.5 GHz wide fast Fourier transform spectrometers, with a spectral resolution of 38.15 kHz, provide the whole coverage of the Q-band in each polarization. The velocity resolution was smoothed to 4$\sim$5 km s$^{-1}$ to improve the signal-to-noise ratio (SNR). In this paper, the observing results of NH$_3$ (13, 13), (14, 14), (15, 15), (17, 17), (18, 18) were presented. The results of other molecular lines will be presented in a future paper. 

We performed OTF mapping observations of molecular lines at 7-mm wavelength toward G0.66-0.13 in the Galactic center. We first mapped an area of $7' \times 7'$, centered on 17:47:46.5, -28:24:47 (J2000). The off position is 17:46:30.3, -28:17:25~\citep{Belloche2013}. Based on the result, deep integration observations were made toward 17:47:44.2, -28:25:04.0 (J2000) (referred as (a)) and 17:47:49.3, -28:23:32.0(J2000) (referred as (b)). The System temperatures vary from 120 K at 32 GHz to 600 K at 50 GHz. 

\subsection{TMRT K band observations}
We performed point-by-point mapping observations of NH$_3$ (3, 3) - (7, 7) at K band toward G0.66-0.13 with a grid of 25$''$. The half power beam width is $\sim$ 45$''$. Similar to the Yebes Q band mapping observations, the Tianma K band mapping area is $7' \times 7'$, while the mapping center is 17:47:46.5, -28:24:47 (J2000). Position-switching was used, with off position 0.5$^{\circ}$ away in azimuth direction. Each scan consists of 2 minutes ON and 2 minutes OFF. The digital backend system (DIBAS) mode 2 was used, with frequency bandwidth of 1.4 GHz, and velocity resolution of 1.1 km s$^{-1}$. Two 1.4 GHz banks were employed, covering a frequency range of 23.2-25.8 GHz. The system temperatures range from 100 K to 200 K during the observations. 1 or 2 scans were observed for each position. The elevation was around 20$^{\circ}$ during observations, thus we adopted an aperture efficiency of 0.45~\citep{Wang2017} to obtain $T_A^*$.

\subsection{TMRT Ka band observations}
NH$_3$ (9, 9) was observed toward G0.66-0.13 (a) and (b) with TMRT at Ka band. The DIBAS backend mode 2 was employed, providing a 1.4 GHz bandwidth across the  26.8-28.2 GHz frequency range. The velocity resolution was $\sim$0.99 km s$^{-1}$. It was smoothed to $\sim$4 km s$^{-1}$ to improve the SNR. Position switching observing mode was used, with off position 0.5$^{\circ}$ away in azimuth direction. Each scan consists of 2 minutes ON and 2 minutes OFF. The system temperatures range from 90 K to 300 K during the observations, which varied with frequencies and weather conditions. We adopted an aperture efficiency of 0.5~\citep{Wang2017} to obtain $T_A^*$.

\subsection{Data reductions}
All the data processing was conducted using \textbf{GILDAS} software package\footnote{\tt http://www.iram.fr/IRAMFR/GILDAS.}. Linear baseline subtractions were used for most of the spectra. For each transition, the spectra of subscans, including two polarizations, were averaged to reduce rms noise levels. Gaussian fitting is used to derive the physical properties of molecule lines for highly-excited transitions (J$>$ 8), including peak intensity, V$_{LSR}$, FWHM line width, and integrated intensity. The results toward G0.66-0.13(a) and (b) are summarized in Table~\ref{tab:g0.66a} and \ref{tab:g0.66b}. As some lowly-excited lines have multiple peaks, we used task `print moment' in GILDAS/CLASS to obtain the physical properties for lines with J$<$8 .

\section{Results}
\label{results}

\subsection{Detection of NH$_3$ highly-excited metastable inversion lines}

\begin{figure}
\begin{center}
\includegraphics[width=0.95\textwidth]{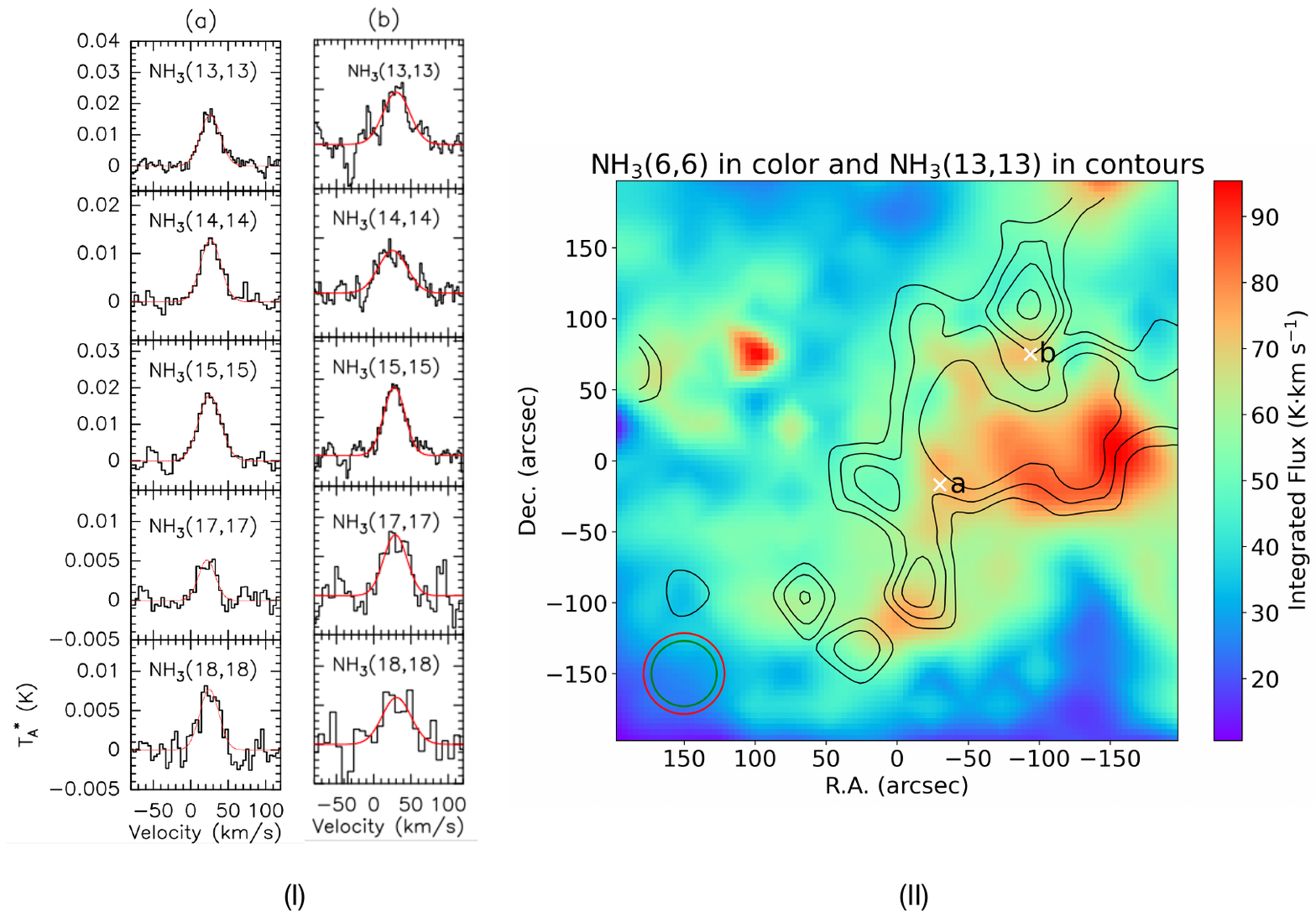}
\caption{(I), The spectra of NH$_3$ (13, 13) - (18, 18) toward positions (a) and (b) in G0.66-0.13. The black lines represent the observations, while the red lines represent the Gaussian fitting result. These spectra were observed with the Yebes 40m telescope. (II), The spatial distribution of NH$_3$ (13, 13) (contours) overlayed on that of NH$_3$ (6,6) (color-scale) toward G0.66-0.13. NH$_3$ (13,13) was mapped with the Yebes 40 m telescope, while the NH$_3$ (6,6) is mapped with TMRT. The contours represent 5$\sigma$ to 17$\sigma$ in steps of 3$\sigma$, with 1$\sigma$ of 0.075 K km s$^{-1}$. The white crosses denote the positions toward which long integration time observations were carried out and rotation diagrams were constructed. The green and red circles in the left corner denote the beamsizes for observations of NH$_3$ (6, 6) and (13, 13), respectively. }
\label{fig:map} 
\end{center}
\end{figure}

From the deep integration observations toward G0.66-0.13 with the Yebes 40m telescope, the metastable NH$_3$ (13, 13), (14, 14), (15, 15), (17, 17) and (18, 18) lines were detected (see Figure~\ref{fig:map}-(I)). Note that the NH$_3$ (16, 16) (39940.95 MHz) is blended with strong emission from HC$_5$N (39939.92 MHz), so the NH$_3$ (16,16) cannot be used. The parameters of the observed highly-excited lines of NH$_3$ toward G0.66-0.13 (a) and (b) are presented in Table~\ref{tab:g0.66a} and \ref{tab:g0.66b}. Notably, we detected NH$_3$ (17, 17) and (18, 18) lines in emission for the first time in interstellar space, with upper energy levels ($E_u$) of 2800 K and 3100 K, respectively (see Figure~\ref{fig:map}-(I)). NH$_3$ (18, 18) had previously only been detected in absorption toward Sgr B2(M)~\citep{Wilson2006}. As no star-forming activity has been found in G0.66-0.13, the results present here imply that the (18, 18) line arises from the extended envelope of Sgr B2, thereby confirming the existence of an extended hot molecular gas component in the CMZ. 
  
\subsection{Spatial distribution of NH$_3$ (6, 6) and (13, 13) }\label{result:spatial} 

NH$_3$ (3, 3) to (7, 7) lines were seen with high SNR toward majority part of the region with the TMRT. The rms noise level range from 0.02 K to 0.04 K under velocity resolution of $\sim$4.3 km s$^{-1}$ (T$_A^*$). The LSR velocities range from -10 to 120 km s$^{-1}$. 

The spatial distribution of NH$_3$ (13,13) line was used to represent the spatial distribution of highly-excited lines. Figure~\ref{fig:map}-(II) presents spatial distribution of the (13, 13) overlayed on that of NH$_3$ (6, 6) in color scale. The spatial distribution of NH$_3$ (13, 13) differs from that of NH$_3$ (6, 6), as its peak offsets the peak emission of NH$_3$ (6, 6). These results imply distinct spatial distributions of warm ($\sim100$~K) and hot ($>400$~K) gas, as represented by NH$_3$ (6, 6) and (13, 13) lines, respectively. The warm gas exhibits a more spatially and spectrally extended, and intense distribution compared to the hot gas. 

\subsection{Estimate of physical conditions}

We performed a rotational diagram analysis to derive the rotational temperatures ($T_{rot}$) and column densities of NH$_3$, in which $T_{rot}$ is close to but less than the actual kinetic temperature~\citep{Morris1973, Danby1988}. The rotational diagram analysis consists of a plot of the column density per statistical weight, for a given number of molecular rotational energy levels, as a function of their energies above the ground state. Under the assumption of local thermodynamic equilibrium (LTE), the level population can be represented by the Boltzmann distribution and the rotation diagram is described by equation (2)~\citep{Goldsmith1999}:
\begin{equation}
ln\frac{N_u}{g_u} = ln\frac{3k_B W}{8\pi^3 \nu S_{ul}\mu^2 g_u} = ln\frac{N}{Q} - \frac{E_u}{k_B T_{ex}}
\end{equation}
in which $N_u$ is the column density of the upper level, $g_u$ is the degeneracy of the upper level, $k_B$ is the Boltzmann constant, $W$ is the integrated intensity of the transition, $\nu$ is the rest frequency of the line, $S_{ul}$ is the line strength of the transition, $\mu$ is the internal partition function of the molecule, and $E_u$ is the upper level energy of the transition. As the ortho/para ratio of NH$_3$ in CMZ has been found to range from 1.0 to 1.3~\citep{Mills2013}, we adopted a ratio of 1.0 for simplicity, which will not alter the conclusion of this work. 

\begin{figure}
\begin{center}
\includegraphics[width=0.95\textwidth]{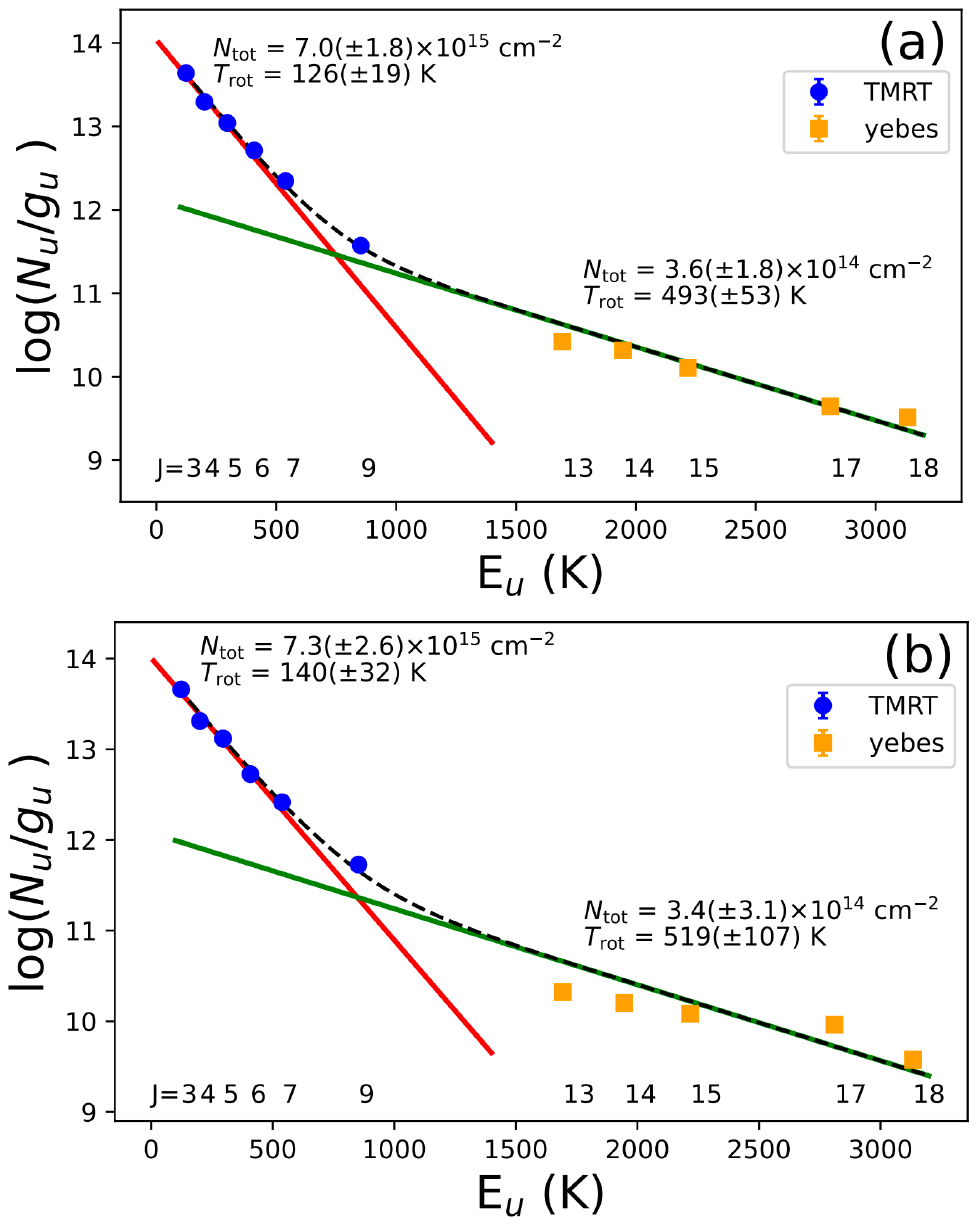}
\caption{Rotational diagrams of NH$_3$ (3, 3) -(18, 18) transitions toward positions (a) and (b) in G0.66-0.13. Blue dots indicate transitions observed with TMRT, while orange squares indicate transitions observed with the Yebes 40m telescope.
\label{fig:rotation}}
\end{center}
\end{figure}

Two temperature components were adopted to account for both the lowly-excited and highly-excited lines. A dominant warm component ($T_{rot}= 120-140$ K) traced by low $E_u$ transitions, and a minor hot component ($T_{rot} >400$ K) traced by high $E_u$ lines were obtained (see Figure~\ref{fig:rotation}). The column density of hot NH$_3$ is $\sim3.5\times10^{14}$ cm$^{-2}$, which amounts to $\sim5$\% of the total NH$_3$, suggesting that only a small fraction of molecular gas is subject to intense heating processes. Our analysis accounts for beam dilution effects by assuming uniform emission over scales larger than beam size~\citep{Mills2013}. If the emission arises from compact sources, corrections to the observed brightness temperatures would be required due to varying beam sizes at different frequencies (see Figure~\ref{fig:rotate_c}). 

 In Figure \ref{fig:rotation}, the rotation temperatures are derived by assuming that NH$_3$ is extended compared to the telescope beam~\citep{Mills2013}. As the molecular lines come from different telescopes, the Shanghai 65m and Yebes 40m, and different frequencies, the beam sizes vary from $37''$ to $57''$ (see Table~\ref{tab:line}). The beam dilution effect on the estimation of temperatures should be investigated. The brightness temperature ($T_B$) can be derived from the main beam brightness temperature ($T_{mb}$) dilution:
 \begin{equation}
 T_B=T_{mb}/\eta_{BD} = T_{mb} \frac{\theta_s ^2 +\theta_{beam}^2}{\theta_s^2},
 \end{equation}
 in which $\eta_{BD}$ is the beam filling factor, $\theta_s$ and $\theta_{beam}$ are source size and the Half-Power Beamwidth (hereafter HPBW), respectively. We adopted a main beam efficiency of 0.6 to convert $T_{A}^*$ to $T_{mb}$ for the TMRT data. For the Yebes data, main beam efficiency of 0.62 was used for NH$_3$ (13, 13) and (14, 14), while 0.6, 0.56, 0.53 were used for NH$_3$ (15, 15), (17, 17), and (18, 18), respectively\footnote{\tt https://rt40m.oan.es/rt40m\_en.php.}. If the molecular line emission comes from compact hot cores, the source size should be several arcseconds~\citep{Bonfand2017}. The source size was assumed to be 5$''$ to derive $T_B$. The rotational diagrams with beam dilution effect correction are shown in Figure~\ref{fig:rotate_c}. We obtained temperatures above 400~K even for point-like sources. The derived column densities are about 2 orders of magnitude greater than those derived with the assumption of extended distribution. The beam dilution effect does not change the conclusion of the work.

\section{Discussion}
\label{discussion}

\subsection{Line cooling timescale of hot molecular gas}

The characteristic cooling timescale of the interstellar gas is important for heating mechanism analysis. The gas thermal energy per H nucleus is 
\begin{equation}
e=\frac{1}{2}(\frac{3}{2}k_B T) = 10^{-15} (\frac{T}{10~K})~erg 
\end{equation}
for a monatomic gas and H$_2$ acts like a monatomic gas at low temperature because its rotational degrees of freedom cannot be excited. The factor of 1/2 comes from 2 H nuclei per H$_2$ molecule~\citep{Krumholz2011}. 

For a gas number density range of $10^4$ to $10^6~cm^{-3}$, the line cooling via rotational lines of CO, its isotopologues $^{13}$CO and C$^{18}$O, and other species~\citep{Goldsmith2001, Ao2013}
\begin{equation}
\Lambda_{gas} = 6 \times 10^{-29} n^{1/2} T_{kin}^{3} dv/dr~erg~cm^{-3}s^{-1}
\label{linecooling}
\end{equation}
in which d$v$/d$r$ is the velocity gradient in km~s$^-1$~pc$^{-1}$. Taking $T= 500~K$, $n=10^{4.5}$cm$^{-3}$  and $dv/dr=3.5$ km~s$^-1$~pc$^{-1}$, one finds a cooling rate of $4.7\times10^{-18}~erg~cm^{-3}~s^{-1}$. The corresponding cooling time is $\sim$7 yr. Even though only $\sim5$\% of the total gas are hot gas with a temperature above 400 K, it is hard to sustain such amount of this hot molecular gas as hot phase given its rapid cooling timescale. 

\subsection{Heating mechanism of hot molecular gas}

\begin{figure}
\begin{center}
\includegraphics[width=0.95\textwidth]{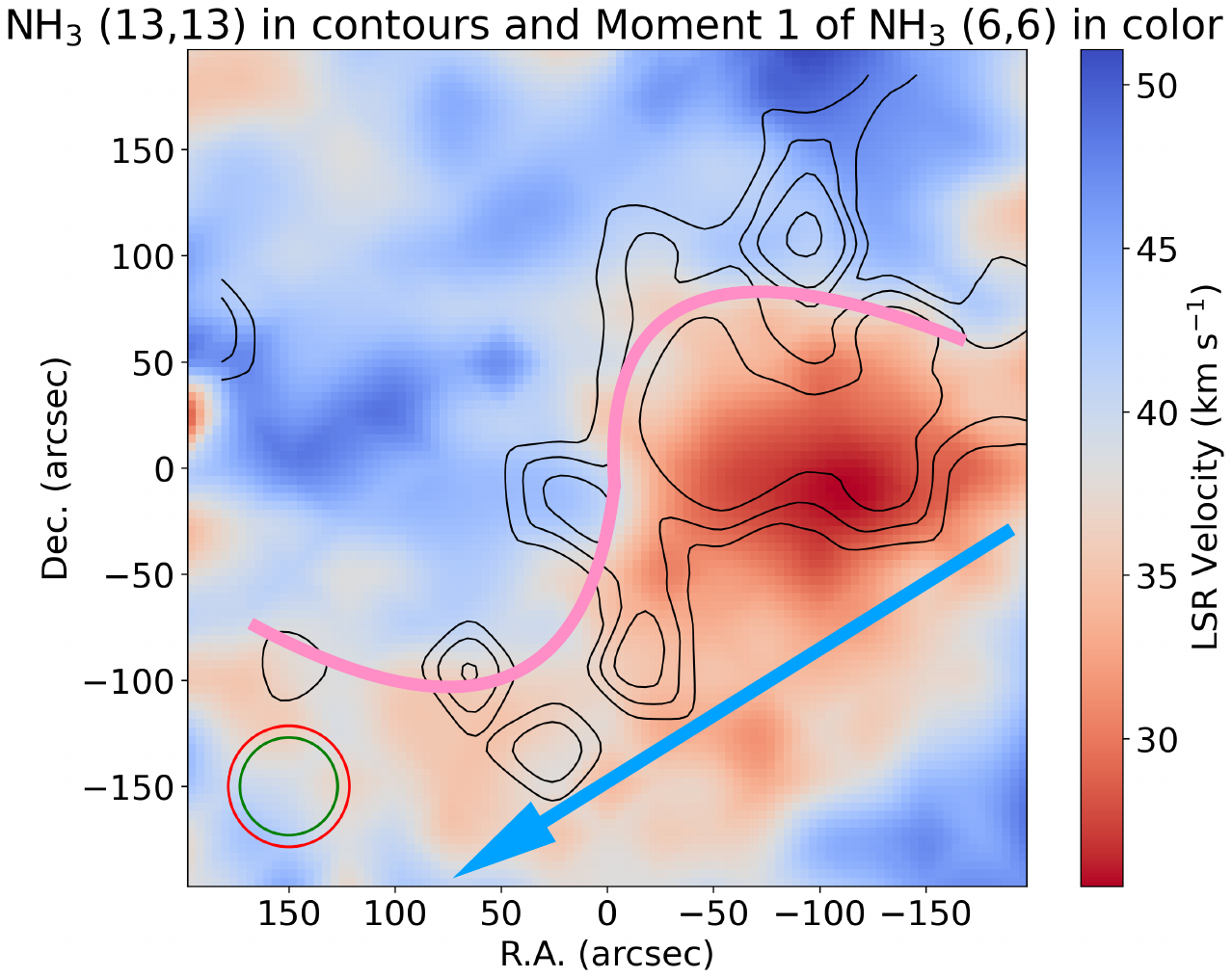}
\caption{The distribution of integrated intensity of NH$_3$ (13, 13) (contours) overlayed on intensity-weighted mean velocity map of NH$_3$ (6, 6) in color scale. FThe contours represent 5$\sigma$ to 17$\sigma$ in step of 3$\sigma$, with 1$\sigma$ of 0.075 K km s$^{-1}$. The blue line represents the orbit of stream around the gravitational potential center of the Galaxy~\cite{Kruijssen2015}. The pink curves mark the boundaries of different gas components. The green and red circles in the left corner denote the beamsizes for observations of NH$_3$ (6, 6) and (13, 13), respectively. 
\label{fig:moment}}
\end{center}
\end{figure}

To explore the origin of this hot molecular gas, we compared the distribution of NH$_3$ (13, 13) with the intensity-weighted velocity map (moment 1 map) of NH$_3$ (6, 6) (see Figure~\ref{fig:moment}). The moment 1 map reveals two main velocity components, with LSR velocities ranging from 25 to 50 km s$^{-1}$. The northeastern region shows a redshifted component centered at 45 km s$^{-1}$, while the southwestern clump exhibits a blueshifted component at 30 km s$^{-1}$. A sharp velocity boundary is found between the northeastern redshifted component and the southwestern blueshifted gas. The hot gas traced by NH$_3$ (13, 13) concentrates along the boundary. The observed velocity boundary aligns spatially with the G0.66-0.13 orbital trajectory derived from CMZ dynamical models~\citep{Kruijssen2015}, indicating shear motion at the interface during the molecular clouds' orbital motion around the Galactic center. %We also compare the line profiles of NH$_3$ (6, 6) between the sheared region and the non-sheared region. As shown in Figure~\ref{fig:moment}-(b), the FWHM line width of lines toward the sheared region is up to 55 km~s$^{-1}$, which is obviously broader than those toward the non-sheared region. These results indicate enhanced turbulence in the sheared region. 

The most common mechanisms for heating the gas in molecular clouds include X-ray heating, cosmic-ray heating, and turbulent heating. Previous quantitive studies have excluded X-ray heating and cosmic-ray heating as the dominant heating mechanisms for warm gas (60-100 K) in the GC, as both the observed X-ray flux density and cosmic ionization rates are too low to explain the heating of the warm gas~\citep{Ginsburg2016, Ao2013}. In addition, both the X-ray heating and cosmic-ray heating are homogenous, thus they could not explain the complex temperature structure observed in the CMZ. Turbulent dissipation has emerged as the most promising heating mechanism for warm gas in the GC~\citep{Ginsburg2016, Ao2013}. Early studies employed spatially averaged viscous dissipation rates to model the turbulent heating~\citep{Ao2013}. In this way, turbulent heating could well explain heating of the warm gas component, which occupies $\sim$95\% of the total gas, but it fails to explain heating of the hot gas component. 

We demonstrate that the intermittency turbulence theory - first proposed by Kolmogorov and Obukhov~\citep{Kolmogorov1962, Obukhov1962} - provides a more complete framework for understanding the CMZ's complex temperature structure. The key insights come from recognizing that turbulent energy dissipation occurs predominantly in localized structures (vortex tubes and shocks) occupying $<1\%$ of the volume, creating extreme temperature variations. Modern magnetohydrodynamic simulations confirm this intermittency, showing that dissipation rate follows a log-Poisson distribution~\citep{She1994, She1995, Pan2009}. 

\begin{figure}
\begin{center}
\includegraphics[width=0.95\textwidth]{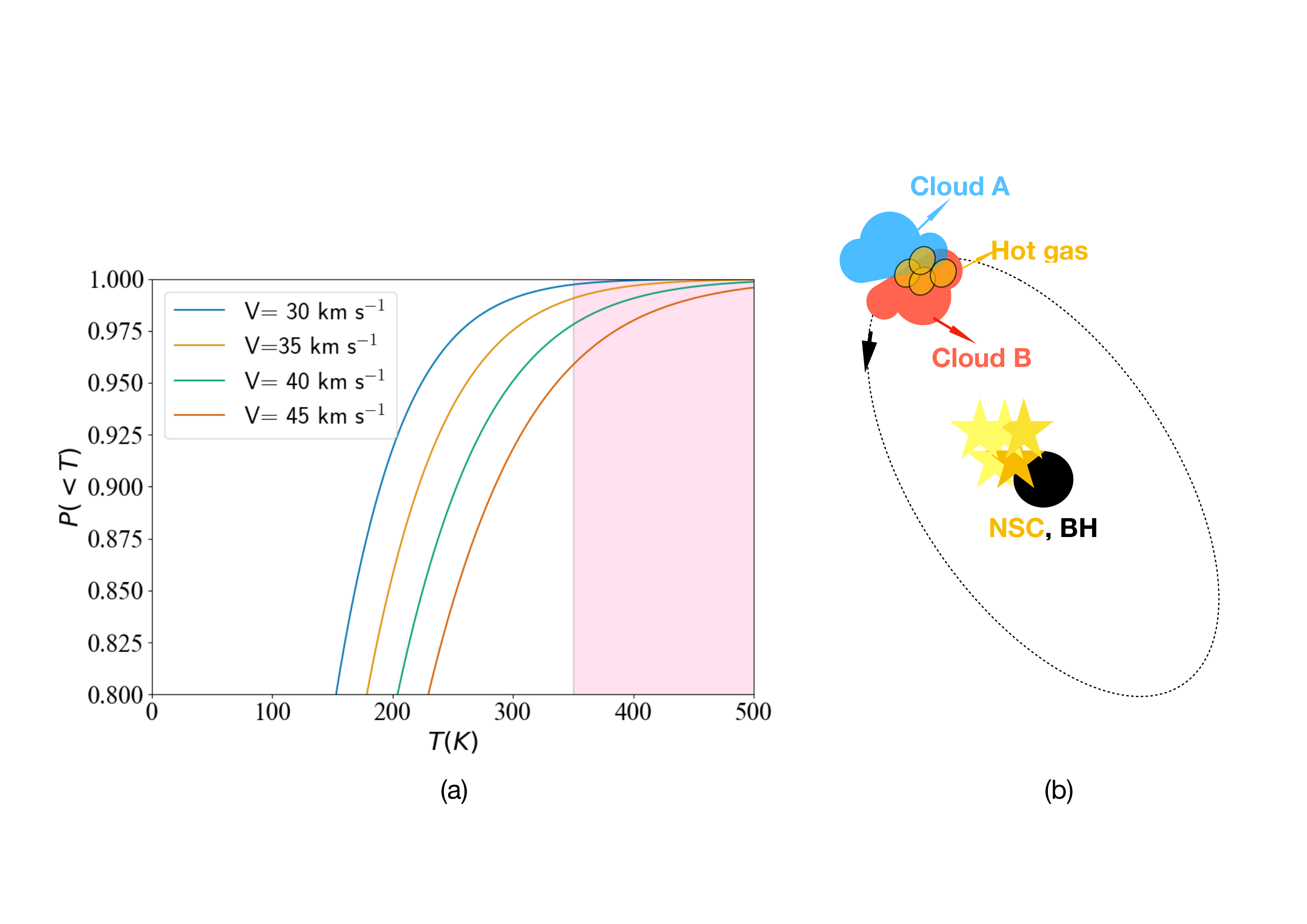}
\caption{(a), Cumulative probability of temperature smaller than $T$ in a molecular cloud with different line widths. The results are derived by assuming that the dissipation rate follows log-Poisson distribution, which is consistent with the intermittency turbulence model. The cloud size was assumed to be 5 pc. A pink box is used to highlight the high-temperature fraction predicted by the model. (b), A schematic picture of the proposed scenario on how hot molecular gas was generated in the Galactic center. Cloud A and Cloud B rotate around the Galactic center under the gravitational potential of the NSCs and the supper massive black hole. 
\label{fig:schematic}}
\end{center}
\end{figure}

\citet{Pan2009} developed an extension of the log-Poisson intermittency model to supersonic turbulence (see detail in the APPENDIX). They computed the correlation coefficient of the dissipation rate and the gas density, obtaining a correlation coefficient of -0.11. Consequently, the dissipation rate and the gas density was assumed to be uncorrelated in their model. We use this intermittency model to compute the probability distribution of the gas temperature for G0.66-0.13 cloud (Figure~\ref{fig:schematic}-(a)): for gas with 40 km s$^{-1}$ line widths, 5 pc cloud size, and 3.5 km s$^{-1}$  pc$^{-1}$ velocity gradient, the intermittent dissipation produces temperatures $>$ 400 K for several percent of the volume. For the wam gas component, the model predicts that P(T$< $350 K)= 0.975 while adopting 40 km s$^{-1}$ line widths, 5 pc cloud size, and 3.5 km s$^{-1}$  pc$^{-1}$ velocity gradient. This means that the warm gas component accounts for over ninety percent of total molecular gas, which is consistent with the relative column densities of the warm component obtained with the rotational diagram analysis. 

Crucially, higher velocity dispersions systematically increase the hot gas fraction and average gas temperatures. According to Equation~\ref{averageT}, for gas with 5 pc cloud size, and 3.5 km s$^{-1}$  pc$^{-1}$ velocity gradient, the average temperatures are 105 K, 122.5 K, 140 K, 157.5 K  for line widths of 30 km s$^{-1}$, 35 km s$^{-1}$, 40 km s$^{-1}$, 45 km s$^{-1}$, respectively. The cloud size is also a critical parameter. If the G0.66-0.13 cloud consists of small clumps with angular size smaller than 40$''$, it will remain unresolved because of the coarse resolution. By adopting a cloud size of 1~pc (corresponding to 25$''$), a line width of 25 km s$^{-1}$ could produce quite a lot of hot gas (see Figure~\ref{fig:cloudsize1pc}). With cloud size of 1~pc, and velocity gradient of 3.5 km s$^{-1}$  pc$^{-1}$, the average gas temperatures are 119 K, 138 K, 159 K, 179 K for line widths of 20 km s$^{-1}$, 25 km s$^{-1}$, 30 km s$^{-1}$, 35 km s$^{-1}$, respectively. Smaller cloud size will systematically increase the hot gas fraction, which could well explain previous detection of hot gas in the CMZ other than the Sgr B2 cloud complex~\citep{Mills2013}.

Turbulence decays rapidly, necessitating a continuous energy source to maintain the observed high turbulence levels in the CMZ~\citep{MacLow1998}. Recent hydrodynamics simulations demonstrate that shear forces arising from differential motion of molecular clouds can effectively convert orbital kinetic energy into turbulent energy, which ultimately dissipates as heat~\citep{Kruijssen2019, Petkova2023}. Our results provide observational evidence supporting that shear-induced turbulence is the promising heating mechanism for molecular gas in the CMZ. The process seems to occur as follows: molecular clouds orbit the Galactic center under the combined gravitational influence of the nuclear stellar cluster (NSCs) and the supermassive black hole~\citep{Schodel2014, gravity2020}. The velocity gradient between adjacent clouds generates shear motions which convert orbital kinetic energy into turbulent energy. The turbulent energy subsequently dissipates, heating a portion of molecular gas to temperatures exceeding 400 K. This complete energy conversion pathway is illustrated schematically in Figure~\ref{fig:schematic}-(b). This scenario may represent a universal mechanism in galactic nuclei. The physical conditions of the CMZ show remarkable similarity to those in NGC 253, IC 342, and Maffei 2~\citep{Henkel2000, Meier2015, Ott2005, Gorski2017, Gorski2018}, suggesting that continuous shear-induced turbulence heating could be a ubiquitous factor influencing galactic evolution.

The molecular clouds in the CMZ, such as Sgr A  complex, Sgr B2, Sgr C, and Brick, are characterized by strong SiO emission, complex temperature structures~\citep{Huettemeister1993, Ao2013, Mills2013, Ginsburg2016}, abundant complex organic molecules~\citep{Requena2006, Requena2008}, low star-formation efficiency, more turbulent than disk clouds and so on~\citep{Bally1987, Henshaw2023}. G0.66-0.13 exhibits all the properties that are common to molecular clouds in the CMZ~\citep{Tsuboi2015, Armijos2020, Banda2023, Li2017, Li2020}, making it a representative example of non-star-forming regions in the CMZ. The total mass of molecular gas within R$\sim$ 300~pc is estimated to be $2-6\times10^7~M_{\odot}$~\citep{Dahmen1998}. Given that the star-forming rate of the CMZ is only $\sim~0.07~M_{\odot}$~yr$^{-1}$~\citep{Lu2019, Henshaw2023}, most of the gas is highly turbulent and lacking star-forming activity~\citep{Henshaw2023}. Thus we could say that the shear-induced turbulence heating should be applicable to the majority of molecular gas in the Galactic center.

\section{Conclusions}
\label{conclusion}

We investigated the physical environments of the CMZ through multi-levels NH$_3$ metastable inversion line observations from (3, 3) to (18, 18) toward G0.66-0.13 with the Yebes 40m telescope and Shanghai Tianma 65m radio telescope.  The main results of this work include:

1. The NH$_3$ (17, 17) and (18, 18) emission lines were detected for the first time in the interstellar medium, which confirms the widespread presence of hot molecular gas ($>$400 K) in the Galactic center. Rotational diagram analysis indicate that the hot NH$_3$ amounts to about 5\% of total NH$_3$.

2. We found spatial separation between the hot and warm gas, which is represented by NH$_3$ (6, 6) and (13,13). The hot gas was found to be concentrated at shear boundary of different cloud components. 

3. The intermittency turbulence heating theory, in which the dissipation rate follows a log-poisson distribution, could well produce the CMZ's complex temperature structures. The turbulence is likely induced by shear-motion of molecular clouds around the gravitational potential center. 

Being the closest galactic nucleus, the GC gives us an opportunity to observe processes that potentially have wide applicability in centers of other galaxies. Our results demonstrate that the intermittent turbulence model elevates the hot molecular gas in the GC to exceed 400 K. Future ALMA and VLA observations can map energy dissipation rates through observations of high-excitation NH$_3$ lines, and further make a judgement of the intermittency turbulence model. These observations will provide critical insights into the nature of turbulence in the center of the Milky Way~\citep{Kruijssen2015, Ginsburg2016}, as well as other normal spiral galaxies.

\begin{acknowledgements}
{We thank the anonymous referee for their comments, which helped us improve this work.} We thank Dr. Paul Ho, Xi Chen and Guangxing Li for the very helpful discussions and constructive suggestions. We are grateful to the Shanghai Tianma 65m and Yebes 40 m telescopes staff for their help during the observing runs. The 40 m radio telescope at Yebes Observatory is operated by the Spanish Geographic Institute (IGN, Ministerio de Transportes, Movilidad y Agenda Urbana). The MeerKAT telescope is operated by the South African Radio Astronomy Observatory, which is a facility of the National Research Foundation, an agency of the Department of Science and Innovation. The ATLASGAL project is a collaboration between the Max-Planck-Gesellschaft, the European Southern Observatory (ESO) and the Universidad de Chile. It includes projects E-181.C-0885, E-078.F-9040(A), M-079.C-9501(A), M-081.C-9501(A) plus Chilean data. This work has been supported by the National Key R\&D Program of China (No. 2022YFA1603101). This work is supported by State Key Laboratory of Radio Astronomy and Technology. AVG acknowledges support from the Spanish grant PID2022-138560NB-I00, funded by MCIN/AEI/10.13039/501100011033/FEDER, EU. L. H. is supported by the Key Program of the National Natural Science Foundation of China (grant no. 11933007, 12325302); the Key Research Program of Frontier Sciences, CAS (grant no. ZDBS-LY-SLH011); the Shanghai Pilot Program for Basic Research, Chinese Academy of Sciences, Shanghai Branch (JCYJ-SHFY-2022-013). 
\end{acknowledgements}

\vspace{5mm}

\bibliography{nh3}{}
\bibliographystyle{aasjournal}

\begin{table*}[htbp!]
      \begin{tabular}{cccccc}
    \hline
    \hline
%              &                  &                                        \\
Molecule & Transition &    Rest Freq.   &   $E_u/k_b$         &   Beam Size        &      Telescope           \\
              &   &   (GHz)           &     (K)                &   ($''$)          &                 \\
\hline        
              &  (1,1)       &    23.6945          &    24      &  49        &      Tianma 65m telescope              \\      %    map
              &  (2,2)   &      23.7226        &       65       &   49         &     Tianma 65m telescope         \\         %  peak:   0.3 K map
              &   (3,3)  &    23.8701        &    124         &    49             &       Tianma 65m telescope      \\      %    
              &  (4,4)    &   24.1394         &     201      &  48           &        Tianma 65m telescope                 \\      %    map
NH$_3$   &  (5,5)   &   24.5329         &   296        &   46           &       Tianma 65m telescope       \\         %  peak:   0.3 K map
              &   (6,6)  &    25.0560         &   409        &   46            &        Tianma 65m telescope        \\     
              &   (7,7)       &  25.7152     &    539       &    45               &     Tianma 65m telescope               \\      %    map
              &   (9,9)       &   27.477943    &  853    &     42               &     Tianma 65m telescope               \\      %    map                               
  %            &   (12,12)    &  31.4249    &  1456   &     37               &     Tianma 65m telescope               \\      %    map                     
              &  (13,13)   &   33.1568   &    1692   &   57       &     Yebes 40m telescope          \\         %  peak:   0.3 K map
              &   (14,14)  & 35.1343    &  1946   &     54          &       Yebes 40m telescope        \\    
              &    (15,15)  &  37.3851  &  2181     &    50           &      Yebes 40m telescope            \\      %    map
               & (17,17)   &  42.8400   &  2811   &    44        &        Yebes 40m telescope      \\         %  peak:   0.3 K map
                & (18,18)  &  46.1233  &   3133  &     41      &      Yebes 40m telescope            \\                                                             
\hline
      \end{tabular}
        \caption{Observed Transitions of NH$_3$.}\label{tab:line}
\end{table*}

\clearpage

\clearpage

\clearpage

%ffffffffffffffffffffffffffffffffff
%ffffffffffffffffffffffffffffffffff

\clearpage

%ffffffffffffffffffffffffffffffffff
%ffffffffffffffffffffffffffffffffff

\clearpage

%ffffffffffffffffffffffffffffffffff
%ffffffffffffffffffffffffffffffffff

\clearpage

\newpage
\appendix

\section{Intermittency turbulence model}
\label{sec:model}

In this section, we describe the intermittency turbulence model that was used to predict the probability distribution of kinematic temperatures (see Fig~\ref{fig:schematic} -(a)). According to numerical simulations, turbulent dissipation is characterized by strong spatial roughness~\citep{Pan2009}. Extreme dissipation events occupy a tiny volume or mass fraction, which implies a broad probability distribution of the dissipation rate. The intermittency turbulence theory that describes this phenomenon, was first developed in~\citet{Kolmogorov1962} and \citet{Obukhov1962}. To account for the fluctuations in the dissipation rate, the dissipation rate, $\epsilon_l$, per unit mass at a scale $l$ was defined as:
\begin{equation}
\epsilon_l(x,t) = \frac{1}{\rho_l(x, t)V(l)} \int_{\vert{x}\vert<l} \rho (x+x', t) \epsilon(x+x', t)dx'
\end{equation}
where $V(l)= 4\pi l^3/3$ is the volume of a spherical region of size $l$ and $\rho_l(x, t) = \frac{1}{V(l)} \int_{\vert{x}\vert<l} \rho (x+x', t)dx'$ is the average density of that spherical region. $\epsilon(x+x',t)$ is the function form of turbulent dissipation rate. 

The probability distribution, $P(\epsilon_l)$, of the dissipation rate, $\epsilon_l$, at a scale $l$ can be calculated as
\begin{equation}
P(\epsilon_l) = \frac{1}{ \overline{\rm \rho}V} \int \delta (\epsilon_l - \epsilon_l(x, t)) \rho_l (x, t) dx
\end{equation}
where $V$ is the total volume of the system and $\overline{\rm \rho}$ is the overall average density in the flow. 

Numerical simulations of supersonic turbulence find that the log-Poisson distribution for $\epsilon_l$ provides a good fit to the dissipation rate~\citep{She1994, Dubrulle1994, She1995, Pan2009}. 
In the log-Poisson model, the distribution, $P_L$, of the dissipation rate  mainly depends on the forcing of the flow and thus is probably flow dependent:
\begin{equation}
P (\epsilon_{l})d \epsilon_l = \stackrel{\infty} {\Sigma_{n=0} } exp(-\lambda) \frac{\lambda^n}{n!} P_L(ln(\epsilon_l/\overline{\rm \epsilon}) - \gamma ln(L/l) - nln(\beta))d~ln(\epsilon_l/\overline{\rm \epsilon})) 
\label{poisson}
\end{equation}
where $\lambda=\lambda_{Ll}=\gamma ln(L/l)/(1-\beta)$. $\beta$ is related to the fractal dimension of the most intermittent structures, $d$, and the parameter $\gamma$ by $\gamma/(1-\beta) = D-d$, with $D=3$ being the dimension of the system~\citep{She1994, Pan2008}. 
For vortex tubes, the dimension is $d=1$ and for shocks $d=2$. In supersonic turbulence, approximately one-third of the energy is dissipated through dilatational modes, whereas two thirds is dissipated through solenoical modes, leading to the expectation that the parameter $d$ falls within the range of 1 to 2~\citep{Pan2009}. 

The turbulent heating rate per unit volume is given by
\begin{equation}
\Gamma_{turb}= n\mu m_H \epsilon(x,t)
\end{equation}
where $n$, $m_H$ are the number density and the mass of the hydrogen atom, $\mu$ is the mean molecular weight, which is 2.35 for molecular clouds. $\epsilon(x,t)$ is the turbulent dissipation rate per unit mass. As pointed above, the heating rate is spatially inhomogeneous due to be intermittency of the turbulent dissipation, which resulted to temperature fluctuations. We will use the distribution of $\epsilon$ that is given by Equation \ref{poisson} , and calculate the mass-weighted average temperature and the temperature probability distribution in molecular clouds in the CMZ. 

The average dissipation rate per unit mass, is given by
\begin{equation}
\overline{\epsilon} = 0.5 \sigma_v^3 /L,
\end{equation}
in which $\sigma_v$ is the one-dimensional velocity dispersion and $L$ the size of the cloud. $\sigma_v$ is related to the observed FWHM line widths by the conversion $\sigma_v = V_{FWHM}/2.355$~\citep{Pan2009}. Then the average turbulent heating rate per unit volume is given by 
\begin{equation}
\overline{\Gamma_{turb}}=0.5 n\mu m_H \sigma_v^3 /L.
\end{equation}

The average gas kinetic temperature, $\overline{T_{kin}}$, can be estimated from the thermal equilibrium
\begin{equation}
\Gamma_{cr} + \overline{\Gamma_{turb}}= \Lambda_{g-d} + \Lambda_{gas}.
\end{equation}
where $\Gamma_{cr}$ is the cosmic heating rate~\citep{vanderTak2000}, $\Lambda_{g-d}$ is the gas cooling via gas-dust interaction~\citep{Tielens2005}, $\Lambda_{gas}$ is the line cooling through rotational lines of CO and its isotopologues, and other species~\citep{Goldsmith1978, Goldsmith2001, Papadopoulos2010, Ao2013}. As cosmic rays cannot explain the observed variation in gas temperatures, so CMZ gas temperatures are not dominated by cosmic ray heating, thus we assume that the turbulent heating dominates the heating process:
\begin{equation}
\overline{\Gamma_{turb}}= \Lambda_{g-d} + \Lambda_{gas}.
\end{equation}

$\Lambda_{g-d}$, the gas cooling via gas-dust interaction is given by~\citep{Tielens2005}
\begin{equation}
\Lambda_{g-d} \sim 4\times 10^{-33} n^2 T_{kin}^{1/2} (T_{kin} - T_d) ~erg~cm^{-3}~s^{-1},
\end{equation}
in which $T_d$ is the dust temperature. By comparing with $\Lambda_{gas}$ given by Equation~\ref{linecooling}, it is found that line cooling dominates the cooling process for gas densities $n<10^5~cm^{-3}$ and the assumed parameters as above.

\begin{figure}[htbp]
 \begin{center}
 \includegraphics[width=0.8\textwidth]{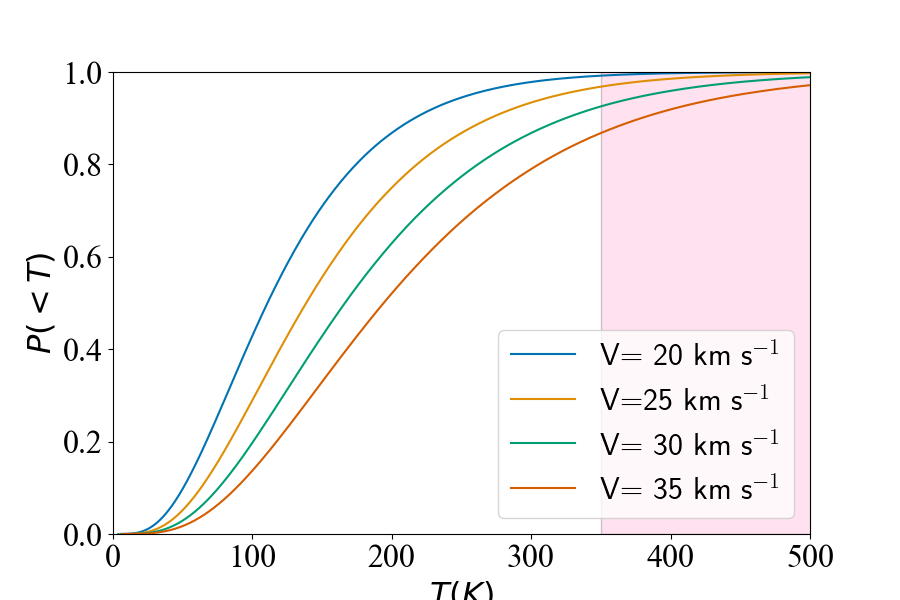}
 \end{center}
 \caption{Cumulative probability of temperature smaller than $T$ in a molecular cloud with different FWHM line widths while adopting a cloud size of 1 pc. The results are derived by assuming that the dissipation rate follows log-Poisson distribution, which is consistent with the intermittency turbulence model. A pink box is used to highlight the high-temperature fraction predicted by the model. }
 \label{fig:cloudsize1pc}
\end{figure}

By assuming that line cooling dominates the cooling process and adopting the observed parameters of a velocity gradient ($3.5$~km~s$^{-1}$~pc$^{-1}$), FWHM line width ($30$~km~s$^{-1}$), and cloud size (5~pc)~\citep{Ao2013}, we derived a $\overline{T_{kin}}$ of $\sim 105$ K for gas densities of $10^{4.5}$ cm$^{-3}$. This means that the average turbulent heating alone could maintain an {average} temperature of $\sim105~K$ in molecular clouds with adopted parameters. In general, the $\overline{T_{kin}} $ can be simplified to~\citep{Ao2013}
\begin{equation}\label{tkin}
\overline{T_{kin}} = 105 ( \frac{n}{10^{4.5} cm^{-3}})^{1/6} ( \frac{L}{5~pc} \frac{dv/dr}{3.5~km~s^{-1} pc^{-1}} )^{-1/3} ( \frac{V_{FWHM}}{30~km~s^{-1}} ) K.
\label{averageT}
\end{equation}
A realistic calculation for the gas temperature needs to take into account the intermittency distribution of the turbulent heating rate. By assuming that the distribution of the turbulent heating rate $\epsilon(x, t)$ follows the log-Poisson distribution, Equation \ref{poisson}, we obtain the cumulative probability of temperature smaller than $T$ in a molecular cloud, which has been shown in Figure~\ref{fig:schematic}-(a). Adopting a cloud size of 1~pc in Equation~\ref{tkin}, Figure~\ref{fig:cloudsize1pc} shows the cumulative probability of temperature smaller than $T$ for different line widths.

\clearpage

%\begin{figure}[htbp]
 %\begin{center}
 %\includegraphics[width=0.95\textwidth]{grid_66.png}
 %\end{center}
 %\caption{Spectra of NH$_3$ (6, 6) emission toward G0.66-0.13 observed with TMRT. The observation was conducted in 2023. For each spectrum the abscissa is the LSR velocity ($V_{LSR}$ = -150 - 200 km s$^{-1}$) and the ordinate is the antenna temperature ($T_{A}^*$ = -0.5 K - 3.5 K). The grid spacing between the spectra is 25$''$. }
 %\label{fig:grid6}
%\end{figure}

%\begin{figure}[htbp]
 %\begin{center}
 %\includegraphics[width=0.95\textwidth]{gridmap_13.png}
% \end{center}
% \caption{Spectra of NH$_3$ (13, 13) emission above 3$\sigma$ level toward G0.66-0.13 observed with the Yebes 40m telescope. The observations were conducted in 2022 and 2024. For each spectrum the abscissa is the LSR velocity ($V_{LSR}$ = -100 - 220 km s$^{-1}$) and the ordinate is the antenna temperature ($T_{A}^*$ = -0.01 K - 0.03 K). The grid spacing between the spectra is 40$''$. }
 %\label{fig:grid13}
%\end{figure}

\begin{figure}[htbp]
 \begin{center}
 \includegraphics[width=0.95\textwidth]{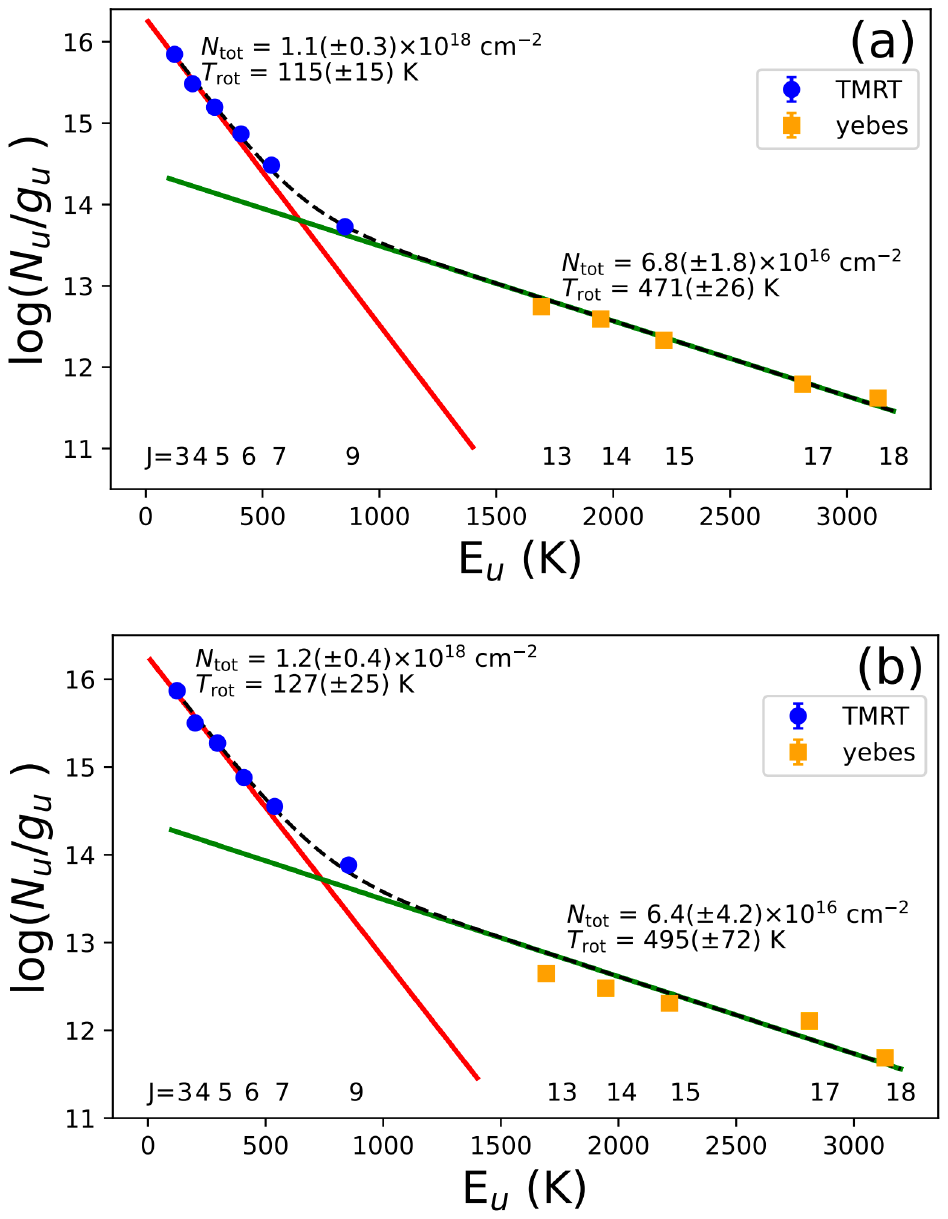}
 \end{center}
 \caption{The rotational diagrams toward G0.666-0.13(a) and (b) after taking into account the beam filling factor. The blue circles denote data observed with TMRT, while the yellow triangles denote data observed with the Yebes 40m telescope.  %\textcolor{red}{(occupy two columns: 183mm, depth: 247 mm, optimum font size is 8pt and no lines thinner than 0.25pt (0.09mm))}
 }
 \label{fig:rotate_c}
\end{figure}

\clearpage

\clearpage

\begin{table*}[htbp]
      \begin{tabular}{ccccc}
    \hline
    \hline
%              &                  &                                        \\
 Transition &    $T_{A}^*$    &   $V_{center}$         &   FWHM               &     Integrated intensity                   \\
                 &   (K)           &     (km s$^{-1}$)       &   (km s$^{-1}$)     &  (K km s$^{-1}$)           \\
\hline   
 %           &        (a)        &   (-30$''$,-17$''$)    &                   &                         \\  % K: east127, Ka: E93               
 (3,3)   &   6.97(0.07)         &  30.6(2.3)     &    47.2(2.3)      &    278.6(0.8)                  \\          
  (4,4)  &   1.96(0.05)        &  31.3(4.5)       &   42.2(4.5)         &    87.1(0.8)                   \\    
   (5,5)  &  1.44(0.05)        &  33.4(4.5)      &    43.5(4.5)        &     62.9(0.8)                              \\       
 (6,6)   &  1.66(0.04)         &   32.5(4.4)     &    41.3(4.4)         &    73.1(0.8)                   \\          
  (7,7)  &   0.50(0.04)        &   29.1(4.3)      &    24.6(4.3)        &     19.0(0.8)                         \\           
 %(9,9)   &   0.33(0.02)         &   37.8(4.0)      &    39.3(4.0)       &   11.9(0.3)                \\  
 (9,9)  &    0.27(0.04)       &   34.5(0.7)   &     30.6(1.8)     &   8.79(0.43)              \\    
%  (12,12)  &   0.17(0.02)      &  36.9(3.5)        &    36.6(3.5)      &    5.6(0.3)                   \\     
%%%%    (12,12)   &  0.09(0.01)     &    34.9(2.0)    &       43.9(4.8)     &    4.2(0.39)                       \\ 
 (13,13)   &   0.016(0.001)         &   26.1(2.8)     &    31.2(2.8)     &    0.54(0.02)                \\         %  peak:   0.3 K map
  (14,14)  &   0.013(0.001)        &  26.5(5.2)      &    32.0(5.2)       &     0.48(0.02)                 \\    
   (15,15)  &  0.018(0.002)        &  25.2(4.9)      &    35.2(4.9)       &     0.67(0.04)                            \\      %    map
 (17,17)   &  0.005(0.001)         &   21.5(4.3)     &    21.8(4.3)       &    0.15(0.01)                  \\         %  peak:   0.3 K map
  (18,18)  &   0.008(0.001)        &   24.8(4.0)     &    30.1(4.0)       &    0.25(0.02)                       \\         
  \hline  
        \end{tabular}
                    \caption{Observing result of NH$_3$ toward G0.66-0.13 (a).}\label{tab:g0.66a}
\end{table*} 

\clearpage

\begin{table}[htbp]
      \begin{tabular}{lccccccccccccc}
    \hline
    \hline
%              &                  &                                        \\
 Transition &    $T_{A}^*$    &   $V_{center}$         &   FWHM               &     Integrated intensity                   \\
                 &   (K)           &     (km s$^{-1}$)       &   (km s$^{-1}$)     &  (K km s$^{-1}$)           \\
\hline     
 %              &      (c)        &     (-94$''$, 75$''$)    &         &                      \\.  %K:   east192, E121,ka   
 (3,3)   &   4.66(0.06)         &   33.4(2.3)     &   57.8(2.3)      &    292.0(0.9)                  \\          
  (4,4)  &   1.5(0.03)        &  33.7(4.5)       &   57.2(4.5)         &    90.4(0.9)                   \\    
   (5,5)  &  1.0(0.03)        &  34.0(4.5)      &    55.8(4.5)        &     75.0(0.9)                              \\       
 (6,6)   &  1.26(0.02)         &   35.2(4.4)     &    54.1(4.4)         &    75.3(0.9)                   \\          
  (7,7)  &   0.41(0.03)        &   37.1(4.3)      &    50.3(4.3)        &     22.2(0.9)                         \\           
% (9,9)   &   0.23(0.03)         &   33.0(4.0)      &    53.2(4.0)       &   14.1(0.3)                \\     
% (9,9)  &  0.24(0.03)    &    24.4(0.7)   &     35.4(1.4)     &     9.0(0.4)       \\   
  (9,9)  &  0.20(0.03)    &    28.9(1.1)   &    58.4(2.8)    &    12.6(0.5)        \\      
%%%%  (12,12)  &   0.12(0.02)          &  31.6(1.4)   &    52.6(3.4)         &  5.68(0.3)                   \\     
   (13,13)   &   0.010(0.002)    &   30.5(2.8)  &  42.6(2.8)      &    0.43(0.03)                  \\          
  (14,14)  &   0.008(0.001)      &  25.5(2.6)   &   45.4(2.6)     &    0.37(0.02)                   \\    
   (15,15)  &  0.018(0.002)       &  27.7(2.4)   &   33.5(2.4)     &    0.64(0.03)                              \\          
  (17,17)  &   0.008(0.002)        &   29.0(4.3)  &    36.9(4.3)   &     0.31(0.03)                         \\           
 (18,18)   &  0.006 (0.002)      &   30.6(7.9)      &    45.3(7.9)    &   0.29(0.05)                \\                                                              
\hline
       \end{tabular}
               \caption{Observing result of NH$_3$ toward G0.66-0.13 (b).}\label{tab:g0.66b}   
\end{table}

\end{document}